%% file: main.tex
\documentclass[conference]{IEEEtran}
\IEEEoverridecommandlockouts
\usepackage[nolist,nohyperlinks]{acronym}
\def\BibTeX{{\rm B\kern-.05em{\sc i\kern-.025em b}\kern-.08em
    T\kern-.1667em\lower.7ex\hbox{E}\kern-.125emX}}

\newcommand*{\mb}[1]{\mathbf{#1}}

\usepackage{enumerate}
\usepackage{algorithmicx, algpseudocode, algorithm}
\algrenewcommand\algorithmicrequire{\textbf{Iteration 0: }}
\algrenewcommand\algorithmicensure{\textbf{Iteration $k$: }}

\usepackage{caption}
\usepackage{graphics}
\usepackage{xparse}
\usepackage{csvsimple}
\usepackage{balance}
\usepackage{ifthen}

\usepackage{hyperref}
\usepackage[nospace]{cite}
\usepackage{amsmath,amssymb,amsfonts}
\usepackage{graphicx}
\usepackage{textcomp}
\usepackage{xcolor}
\usepackage{soul}
\usepackage{listings}
\usepackage{multirow}
\usepackage[export]{adjustbox}
\usepackage{mathabx}
\usepackage{array}
\usepackage{mdwmath}
\usepackage{mdwtab}
\usepackage{eqparbox}
\usepackage{url}
\usepackage{color}
\usepackage{bm}
\usepackage{booktabs}
\newcommand{\ra}[1]{\renewcommand{\arraystretch}{#1}}
\usepackage[tableposition=top]{caption} 


\floatname{algorithm}{Proposed Algorithm}

\ifCLASSOPTIONcompsoc
 \usepackage[caption=false,font=normalsize,labelfont=sf,textfont=sf]{subfig}
\else
 \usepackage[caption=false,font=footnotesize]{subfig}
\fi
\usepackage{stfloats}

\newcommand{\version}{draft}

\input{changes}

\begin{document}
\def\thetitle{A Robust ADMM-Based Optimization Algorithm For Underwater Acoustic Channel Estimation}
\title{\thetitle}
\author{
\IEEEauthorblockN{
    Tian Tian, \IEEEauthorrefmark{1}\IEEEauthorrefmark{2}
    Agastya Raj \IEEEauthorrefmark{4}
    Bruno Missi Xavier\IEEEauthorrefmark{4}\IEEEauthorrefmark{5}
    Ying Zhang \IEEEauthorrefmark{3}
    Feiyun Wu \IEEEauthorrefmark{1}\IEEEauthorrefmark{2},
    Kunde Yang \IEEEauthorrefmark{1}\IEEEauthorrefmark{2},
}

\IEEEauthorblockA{\IEEEauthorrefmark{1}
    School of Marine Science and Technology, Northwestern Polytechnical University, Xi’an, Shaanxi, China}
\IEEEauthorblockA{\IEEEauthorrefmark{2}
    Key Laboratory of Ocean Acoustics and Sensing (Northwestern Polytechnical University), \\Ministry of Industry and Information Technology, Xi’an, Shaanxi, China.}
\IEEEauthorblockA{\IEEEauthorrefmark{3}
College of Oceanography, Hohai University, Nanjing, Jiangsu, China}
\IEEEauthorblockA{\IEEEauthorrefmark{4}
    Trinity College Dublin, Ireland}
\IEEEauthorblockA{\IEEEauthorrefmark{5}
    Federal Institute of Espírito Santo, Espírito Santo, Brazil}
}
\maketitle

\input{sections/0_abstract}

\input{sections/1_intro}

\input{sections/2_algorithm}
\input{sections/3_simulation}

\input{sections/4_conclusion}
\input{sections/5_acknowledgements}

\input{acronyms}

\bibliographystyle{IEEEtran}
\bibliography{IEEEabrv, Oceans2023}
\end{document}


%% file: changes.tex
\usepackage{xparse}

\ExplSyntaxOn
\NewExpandableDocumentCommand{\ifstringsequalTF}{mmmm}
 {
  \str_if_eq:eeTF { #1 } { #2 } { #3 } { #4 }
 }
\NewExpandableDocumentCommand{\stringcase}{mO{}m}
 {
  \str_case_e:nnF { #1 } { #3 } { #2 }
 }
\ExplSyntaxOff

\ifstringsequalTF{\version}{draft}
  {
    \newcommand{\added}[1]{{\color{blue}#1}}
    \newcommand{\deleted}[1]{{\color{red}\st{#1}}}
    \newcommand{\noted}[1]{{\color{orange}#1}}
  }
  {
    \ifstringsequalTF{\version}{final} {
        \newcommand{\added}[1]{#1}
        \newcommand{\deleted}[1]{}
        \newcommand{\noted}[1]{}
    } 
    {
      \ifstringsequalTF{\version}{old} {
        \newcommand{\added}[1]{}
        \newcommand{\deleted}[1]{{\color{red}#1}}
        \newcommand{\noted}[1]{}
      }
      {
        \newcommand{\added}[1]{{#1}}
        \newcommand{\deleted}[1]{}
        \newcommand{\noted}[1]{}
      }
    }    
  }


%% file: sections/0_abstract.tex
\begin{abstract}
Accurate estimation of the \ac{uwa} is a key part of underwater communications, especially for coherent systems. The severe multipath effects and large delay spreads make the estimation problem large-scale. The non-stationary, non-Gaussian, and impulsive nature of ocean ambient noise poses further obstacles to the design of estimation algorithms. Under the framework of \ac{cs}, this work addresses the issue of robust channel estimation when measurements are contaminated by impulsive noise. A first-order algorithm based on \ac{admm} is proposed. Numerical simulations of time-varying channel estimation are performed to show its improved performance in highly impulsive noise environments. 
\end{abstract}

%% file: sections/1_intro.tex
\section{Introduction}\label{sec:intro}
The discrete input-output relationship of transmitting a signal through \ac{uwa} channel can often be simplified to the following linear expression
\begin{equation}\label{eq: sec2_1}
    \mb{y}=\mb{A x}+\mb{n},
\end{equation}
where $\mb{y}\in\mathbb{C}^M$ is the observation vector of the discrete received signal/symbols, $\mb{x}\in\mathbb{C}^N$ is the vector containing the unknown channel parameters to be estimated, $\mb{A}\in\mathbb{C}^{M\times N}$ is the matrix constructed by training signal/symbols, and $\mb{n}$ is a vector related to noise. The challenge of solving \eqref{eq: sec2_1} in \ac{uwa} environment is that the system is usually underdetermined, which means that the number of training symbols or equations is less than that of unknown parameters. Fortunately, the essentially sparse property of \ac{uwa} makes it possible to solve \eqref{eq: sec2_1} under \ac{cs} framework. One of the most frequently used methods is to convert \eqref{eq: sec2_1} to the $\ell_1$-regularized least squares problem
\begin{equation}   \underset{\mb{x}}{\operatorname{min}}\text{ } ||\mb{y}-\mb{A x}||_2+\lambda|\mb{x}||_1,\label{eq: sec2_l2_cost}
\end{equation}
where $\lambda>0$ is the regularization parameter. However, when the distribution of noise $\mb{n}$ is heavy-tailed or measurements $\mb{y}$ contain impulse interference, the performance of sparse estimation algorithms based on \eqref{eq: sec2_l2_cost} will be degraded. In this case, the cost function
\begin{equation}\label{eq: sec2_l1_cost}
    \underset{\mb{x}}{\operatorname{min}}\text{ } \tau||\mb{y}-\mb{A x}||_1+||\mb{x}||_1
\end{equation}
is more robust because $\ell_1$-norm is less sensitive to outliers compared with using squared error \cite{wenRobustSparseRecovery2017}. Optimizing \eqref{eq: sec2_l1_cost} directly is difficult since both parts are non-differentiable. \ac{admm} provides a feasible framework for solving the above issue \cite{boydDistributedOptimizationStatistical2010a, zhangOnlineProximalADMMTimeVarying2021}. Combined with \ac{pgm} \cite{parikhProximalAlgorithms2014}, an algorithm with high computational efficiency and good impulsive interference resisting ability is developed in this work.


%% file: sections/2_algorithm.tex
\section{The Proposed Algorithm}

\subsection{General Framework}
The optimization problem in \eqref{eq: sec2_l1_cost} can be equivalently reformulated as
\begin{equation}\label{eq: sec3_obj_problem}
    \begin{aligned}
    & \underset{\mb{x,\mb{z}}}{\operatorname{min}}\text{ } \tau||\mb{z}||_1+||\mb{x}||_1\\
    & \text{subject to }\mb{A x+\mb{z}}=\mb{y}
    \end{aligned}
\end{equation}
by introducing an auxiliary variable $\mb{z}=y-\mb{A x}$. Accordingly, the \ac{alf} of \eqref{eq: sec3_obj_problem} is
\begin{equation}\label{eq: sec3_alm_def}
    L_{\rho}(\mb{x},\mb{z},\boldsymbol{\gamma})=||\mb{x}||_1+\tau||\mb{z}||_1+\frac{\rho}{2}||\mb{z}+\mb{A
    x}-\mb{y}+\boldsymbol{\gamma}/\rho||_2^2-\frac{1}{2\rho}||\boldsymbol{\gamma}||_2^2,
\end{equation}
where $\boldsymbol{\gamma}$ is the vector of dual variable associated with equality constraint, and $\rho>0$ is the penalty factor of the \ac{alf}. Under the framework of \ac{admm}, minimization over \ac{alf} is decomposed into two subproblems, and the primary variable $\mb{x}$ and auxiliary variable $\mb{z}$ are updated alternately by
\begin{subequations}
\begin{align}
    & \mb{x}^{(k+1)}=\underset{\mb{x}}{\operatorname{min}}\text{ } f_1(\mb{x})+f_2(\mb{x})\label{eq: sec3_admm_step1}\\
    & \mb{z}^{(k+1)}=\underset{\mb{z}}{\operatorname{min}}\text{ } g_1(\mb{z})+g_2(\mb{z})\label{eq: sec3_admm_step2}\\
    & \boldsymbol{\gamma}^{(k+1)}=\boldsymbol{\gamma}^{(k)}+\rho(\mb{z}^{(k+1)}+\mb{A x}^{(k+1)}-\mb{y})\label{eq: sec3_admm_step3}
\end{align}
\end{subequations}
with
\begin{equation}\label{eq: sec3_def_subfunc}
\begin{aligned}
&f_1(\mb{x})\triangleq\frac{\rho}{2}||\mb{z}^{(k)}+\mb{A x}-\mb{y}+\boldsymbol{\gamma}^{(k)}/\rho||_2^2,\quad
f_2(\mb{x})\triangleq||\mb{x}||_1,\\
&g_1(\mb{z})\triangleq\frac{\rho}{2}||\mb{z}+\mb{A x}^{(k+1)}-\mb{y}+\boldsymbol{\gamma}^{(k)}/\rho||_2^2,\quad g_2(\mb{z})\triangleq\tau||\mb{z}||_1.
\end{aligned}
\end{equation}
One can observe from \eqref{eq: sec3_def_subfunc} that the objective functions for both the $\mb{x}$- and $\mb{z}$-subproblems consist of a smooth, convex, quadratic function and a convex, non-differentiable, $\ell_1$-norm term. Such problems can be solved efficiently by \ac{pgm}. 

The proximal operator is defined as \cite{parikhProximalAlgorithms2014}
\begin{equation}\label{eq: sec3_def_prox}
    {\operatorname{prox}}_{f,t}(x)=\underset{x^+}{\operatorname{argmin}\text{ }}\frac{1}{2t}||x^+-x||_2^2+f(x^+),
\end{equation}
which is also called the proximal mapping of function $f$ with step-size parameter $t$. Comparing \eqref{eq: sec3_admm_step1}, \eqref{eq: sec3_admm_step2} and \eqref{eq: sec3_def_subfunc} with the definition of the proximity operator, one can see that the $\mb{z}$-subproblem conforms to the proximal mapping, while the $\mb{x}$-subproblem does not due to the presence of the matrix $\mb{A}$. One commonly-used strategy to address this issue is to linearize the intractable part of the subproblem \cite{hagerInexactAlternatingDirection2019}. Specifically, for the $\mb{x}$-subproblem here, the differentiable function $f_1(\mb{x})$ can be approximated by its first-order Taylor expansion at the previous iteration's solution $\mb{x}^{(k)}$. Omitting the constant terms related to $\mb{x}^{(k)}$, the approximation form of $f_1(\mb{x})$ can be expressed as
\begin{equation}\label{eq: sec3_quadraApprox_Fx}
\begin{aligned}
\tilde{f}_1(\mb{x})&\triangleq \frac{1}{2t_x}||\mb{x}-(\mb{x}^{(k)}-t_x\nabla f_1(\mb{x}^{(k)}))||_2^2,
\end{aligned}
\end{equation}
where $\nabla f_1(\cdot)$ represents the gradient of function $f_1$.Replacing $f_1(\mb{x})$ with $\tilde{f_1}(\mb{x})$ and substituting it into \eqref{eq: sec3_admm_step1}, both $\mb{x}$- and $\mb{z}$-subproblems can now be solved by \ac{pgm}
\begin{align}
    \mb{x}^{(k+1)}&={\operatorname{prox}}_{f_2,t_x}\left(\mb{x}^{(k)}-t_x\nabla f_1(\mb{x}^{(k)})\right)\label{eq: sec3_prox_xstep}\\
    \mb{z}^{(k+1)}&={\operatorname{prox}}_{g_2,t_z}\left(\mb{z}^{(k)}-t_z\nabla g_1(\mb{z}^{(k)})\right)\label{eq: sec3_prox_zstep}
\end{align}
with gradient vector defined as
\begin{align}
    &\nabla f_1(\mb{x}^{(k)})=\rho\mb{A}^H(\mb{z}^{(k)}+\mb{A x}^{(k)}-\mb{y}+\boldsymbol{\gamma}^{(k)}/\rho)
    \label{eq: sec3_grad_f1}\\
    &\nabla g_1(\mb{z}^{(k)})=\rho(\mb{z}^{(k)}+\mb{A x}^{(k+1)}-\mb{y}+\boldsymbol{\gamma}^{(k)}/\rho).\label{eq: sec3_grad_g1}
\end{align}
Given that $f_2(\mb{x})$ and $g_2(\mb{z})$ both utilize $\ell_1$-norm regularization function, the separable property of $\ell_1$-norm can be leveraged to simplify the evaluation of proximal operators in \eqref{eq: sec3_prox_xstep} and \eqref{eq: sec3_prox_zstep} to one-dimensional minimization problems. The resulting problems can be solved by using the soft-thresholding operator
\begin{equation}
    \mathcal{S}_{\alpha}(\beta)\triangleq \frac{\max(|\beta|-\alpha,0)}{\max(|\beta|-\alpha,0)+\alpha}\beta,
\end{equation}
which leads to
\begin{align}
    \mb{x}^{(k+1)}&=S_{t_x/\rho}\left(\mb{x}^{(k)}-t_x/\rho\cdot\nabla f_1(\mb{x}^{(k)})\right)\label{eq: sec3_xup_mono}\\
    \mb{z}^{(k+1)}&=S_{t_z\tau/\rho}\left(\mb{z}^{(k)}-t_z/\rho\cdot\nabla g_1(\mb{z}^{(k)})
    \right).
\end{align}

\subsection{Step-Size Parameter Setting}
The step-size parameter $t$ in \ac{pgm} is related to the Lipschitz constant (see Sec.4 of \cite{daspremontAccelerationMethods2021}) of the differentiable part of the objective function. Let $L_x$ and $L_z$ denote the Lipschitz constant of function $f_1(\mb{x})$ and $g_1(\mb{\mb{z}})$, respectively. According to \eqref{eq: sec3_def_subfunc}, we have $L_x=\lambda_{\max}({\mb{A}^H\mb{A}})$ and $L_z=1$, where $\lambda_{\max}(\cdot)$ represents the maximum eigenvalue of the given matrix. When the Lipschitz constant is known, the step-size parameter of \ac{pgm} can be set as the reciprocal, i.e., we can assign fixed step-size parameters $t_x=1/\lambda_{\max}(\mb{A}^H\mb{A})$ and $t_z=1$. However, for large-scale problems such as the large delay-spread channel estimation problem in \ac{uwa} environment, the maximum
eigenvalue of $\mb{A}^H\mb{A}$ might be expensive to evaluate. In this case, adopting a simple backtracking line search strategy to
adjust the step-size parameter adaptively is an efficient way
to avoid costly computation of the Lipschitz constant.

\subsection{Residues and Stopping Criteria}
Convergence measures of the \ac{admm} algorithm are derived from primal and dual feasibility conditions, while the residuals of these optimality conditions are often used to define the termination criterion for the \ac{admm} iterations. Boyd et al. gave the definitions of primal and dual residues of the standard \ac{admm} algorithm (see Sec. 3.3 of \cite{boydDistributedOptimizationStatistical2010a}). Instead of minimizing $L_{\rho}(\mb{x},\mb{z}^{(k)},\boldsymbol{\gamma}^{(k)})$, $\mb{x}^{(k+1)}$ here minimizes a linearized \ac{alf} comprised of $\tilde{f}_1(\mb{x})$, as defined in \eqref{eq: sec3_quadraApprox_Fx}. Following the similar derivation process in \cite{boydDistributedOptimizationStatistical2010a}, the primal residue $\mb{r}_p$ and dual residue $\mb{r}_d$ of the proposed algorithm are defined as
\begin{align}
\mb{r}_{p}^{(k+1)}&=\mb{A}\mb{x}^{(k+1)}+\mb{z}^{(k+1)}-\mb{y}\label{eq: sec3_rp}\\
\mb{r}_{d}^{(k+1)}&=\rho\mb{A}^H(\mb{r}_p^{(k+1)}-\mb{r}_p^{(k)})-\frac{1}{t_x^{(k+1)}}(\mb{x}^{(k+1)}-\mb{x}^{(k)})\label{eq: sec3_rd}.
\end{align}
The iteration of the proposed algorithm terminates when conditions
\begin{equation}\label{eq: sec3_stop_ccriteria}
    ||\mb{r}_p^{(k)}||_2\leq\epsilon_p^{(k)}\quad \text{and}\quad ||\mb{r}_d^{(k)}||_2\leq\epsilon_d^{(k)}
\end{equation}
are satisfied, where $\epsilon_p^{(k)}$ and $\epsilon_d^{(k)}$ can update by an abosulte and relative criterion:
\begin{align}
    \epsilon_p^{(k)}&=\sqrt{M}\epsilon_{\text{abs}}+\epsilon_{\text{rel}}\max\{||\mb{A x}^{(k)}||_2,||\mb{z}^{(k)}||_2,||\mb{y}||_2\}\label{eq: sec3_ptol}\\
    \epsilon_d^{(k)}&=\sqrt{N}\epsilon_{\text{abs}}+\epsilon_{\text{rel}}||\mb{A}^{H}\boldsymbol{\gamma}^{(k)}||_2\label{eq: sec3_dtol}.
\end{align}
Above, $\epsilon_{\text{abs}}$ and $\epsilon_{\text{rel}}$ are absolute and relative tolerances respectively, and $M$ and $N$ represent the dimensions of matrix $\mb{A}$ (i.e., $\mb{A}\in\mathbb{C}^{M\times N}$).

\subsection{Penalty Parameter Tuning}
The penalty parameter $\rho$ in \ac{alf} plays an important role in achieving a good convergence rate of \ac{admm} algorithm. A lager value of $\rho$ imposes a larger penalty on violations of equality constraint (see \eqref{eq: sec3_alm_def}), leading to a smaller value of primal residue. Conversely, the definition of dual residue in \eqref{eq: sec3_rd} suggests that a smaller value of $\rho$ contributes to reducing dual residue. To balance these two residuals and ensure their convergence to zero as the iteration proceeds, as well as make the performance of \ac{admm} algorithm less dependent on the initial choice of $\rho$, a simple adjustment scheme that often works well in practice is given by \cite{wohlbergADMMPenaltyParameter2017}
\begin{equation}\label{eq: sec3_rho_update}
    \rho^{(k+1)}=\left\{\begin{array}{ll}
        \delta^{\text{incr}}\rho^{(k)} &  \text{if } ||\mb{r}_p^{(k)}||_2>\xi||\mb{r}_d^{(k)}||_2\\
        \delta^{\text{decr}}\rho^{(k)} & \text{if } ||\mb{r}_d^{(k)}||_2>\xi||\mb{r}_p^{(k)}||_2\\
        \rho^{(k)} & \text{otherwise.}
    \end{array}\right.
\end{equation}
A typical choice of constant parameters in \eqref{eq: sec3_rho_update} is $\xi=10$ and $\delta^{\text{incr}}=\delta^{\text{decr}}=2$.

Finally, the complete pesudocode of the proposed algorithm is summarized in the table below. Here, $J(\mb{x})$ denotes the cost function given in \eqref{eq: sec2_l1_cost}, and $G(\mb{z})=g_1(\mb{z})+g_2(\mb{z})$ corresponds to the objective function defined in \eqref{eq: sec3_admm_step2}.

\begin{algorithm}
\caption{}
\begin{algorithmic}[1]
\State \textbf{Initialization:} 
\State $\quad$ Choose  $\rho^{(0)}$, $\tau$, $\epsilon_{\text{abs}}$, $\epsilon_{\text{rel}}>0$. 
\State $\quad$ Set $\mb{x}^{(0)}=0$, $t_0>0$, and $\eta_x>1$.
\State $\quad$ Set $\mb{z}^{(0)}=\mb{y}-\mb{A x}^{(0)}$.
\State\textbf{for} $k = 1, 2, 3, \dots$ \textbf{do}
\State $\triangleright$ Update the primary variable $\mb{x}^{(k)}$:
\State $\quad$ Compute the gradient $\nabla f_1(\mb{x}^{(k-1)})$ using \eqref{eq: sec3_grad_f1}.
\State $\quad$ Backtracking Line Search: find the minimum number of iterations $l^{(k)}$ such that with $t_x^{(k)}=t_0/\eta^{l^{(k)}}$ and $\mb{x}_{l^{(k)}}=S_{t_x^{(k)}/\rho^{(k)}}\left(\mb{x}^{(k-1)}-t_x^{(k)}/\rho^{(k)}\cdot \nabla f_1(\mb{x}^{(k-1)})\right)$
\[J(\mb{x}_{l^{(k)}})\leq J(\mb{x}^{(k-1)}).\]
\State $\triangleright$ Update the auxiliary variable $\mb{z}^{(k)}$:
\State $\quad$ Compute the gradient $\nabla g_1(\mb{z}^{(k-1)})$ by \eqref{eq: sec3_grad_g1}.
\State $\quad$ Evaluate the proximal operator:  \[\mb{w}^{(k)}=S_{\tau/\rho^{(k)}}\left(\mb{z}^{(k-1)}-\nabla g_1(\mb{z}^{(k-1)})/\rho^{(k)}\right).\]
\State $\quad$ Compute $G(\mb{w}^{(k)})$ and
\[\mb{z}^{(k)}=\underset{\mb{z}}{\operatorname{argmin}}\text{ }\{G(\mb{z}): \mb{z}=\mb{w}^{(k)}, \mb{z}^{(k-1)}\}.\]
\State $\quad$ Update the dual variable $\boldsymbol{\gamma}^{(k)}$ by \eqref{eq: sec3_admm_step3}.
\State $\quad$ Compute the residues $\mb{r}_p^{(k)}$ and $\mb{r}_d^{(k)}$ by \eqref{eq: sec3_rp}-\eqref{eq: sec3_rd}.
\State $\quad$ Calculate the tolerance $\epsilon^{(k)}_p$ and $\epsilon^{(k)}_d$ by \eqref{eq: sec3_ptol}-\eqref{eq: sec3_dtol}.
\State $\quad$ \textbf{if} stopping criterion in \eqref{eq: sec3_stop_ccriteria} is satisfied \textbf{then}
\State $\quad\qquad$ \textbf{break}
\State $\quad$ \textbf{end if}
\State $\quad$ Update the penalty parameter $\rho^{(k)}$ by \eqref{eq: sec3_rho_update}.
\State \textbf{end for}
\end{algorithmic}
\label{alg: proposed}
\end{algorithm}

%% file: sections/3_simulation.tex
\section{Numerical Simulations}
\begin{figure}[!t]
\centering
\subfloat[]{\includegraphics[scale = 0.25]{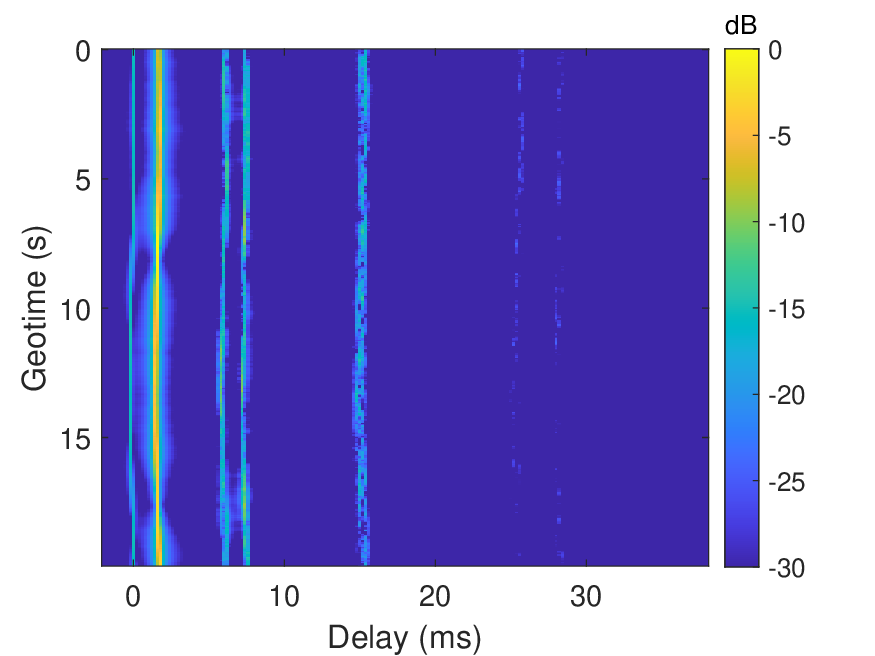}
\label{fig.simu_cir}}
\subfloat[]{\includegraphics[scale = 0.25]{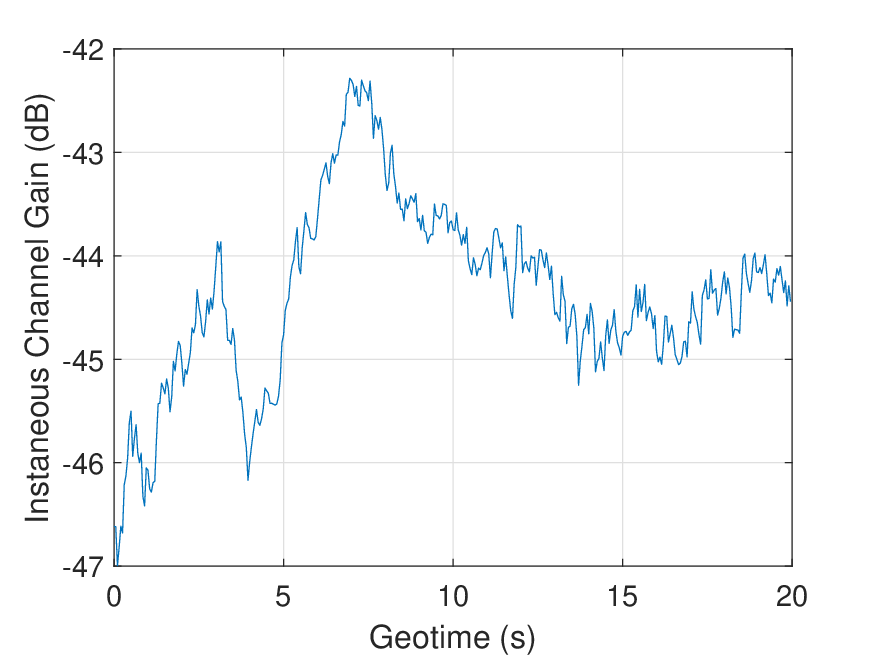}
\label{fig.gain}}
\caption{The simulated time-varying shallow water channel. (a) \ac{cir}. (b) Instantaneous channel gain.}
\label{fig. channel}
\end{figure}

In this section, the performance of the proposed algorithm is tested in solving the sparse channel estimation problem. The PN sequence modulated at baseband with the rate of 5 kbaud of BPSK scheme is used as probe signal. Two first-order methods widely used in engineering: \ac{omp} \cite{caiOrthogonalMatchingPursuit2011} and \ac{fista} \cite{beckFastIterativeShrinkageThresholding2009a} are adopted as comparisons. Samples of time-varying shallow water \ac{cir} are generated by the model introduced in \cite{qarabaqiStatisticalCharacterizationClass2014}. Fig. \ref{fig. channel} shows the simulated \ac{cir} and the variation of instantaneous channel gain. A two-component \ac{gmn} model is applied to simulate the received signal contaminated by impulsive noise:
\begin{equation}
    P(\mb{n}[i])=(1-q)\mathcal{N}(0,\sigma_W^2)+q\mathcal{N}(0,\sigma_I^2),\quad i=1,\cdots,N
\end{equation}
where $q$ represents the probability of occurrence of impulsive noise, $\mathcal{N}(\cdot)$ is the complex Gaussian distribution function, and $\sigma^2_W$ and $\sigma^2_I$ are the variance of the \ac{wgn} and impulsive noise, respectively. Let $\sigma_S^2$ denote the transmitted signal power, the \ac{snr}, \ac{inr}, and \ac{sinr} can be expressed as $\text{SNR}=10\log_{10}(\sigma_S^2/\sigma_W^2)$,  $\text{INR}=10\log_{10}(\sigma_I^2/\sigma_W^2)$, and $\text{SINR}=10\log_{10}(\sigma_S^2/(\sigma_W^2+\sigma_I^2))$, respectively. To evaluate the performance of the proposed algorithms, we conducted simulations under three noise conditions including the \ac{awgn} environment with SNR of 15 dB. For the impulsive noise environment, the \ac{wgn} level is set to the same level as in the \ac{awgn} environment (i.e., with  \ac{snr}$=15$dB), and \ac{inr} is set to 40 dB and 50 dB. The corresponding \ac{sinr} values for these two noise environments are approximately 1.83 dB and -8.90 dB, respectively.

\begin{figure}[!t]
\centering
\subfloat[]{\includegraphics[scale = 0.17]{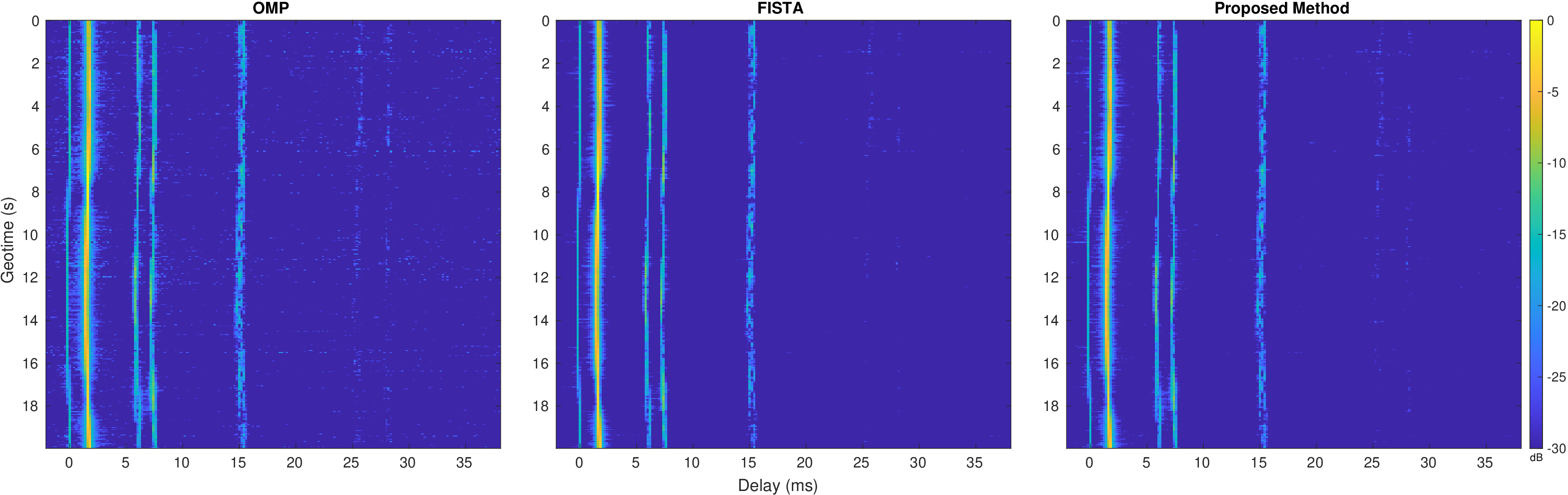}
\label{fig.cir_awgn_snr_15dB}}
\hfill
\subfloat[]{\includegraphics[scale = 0.17]{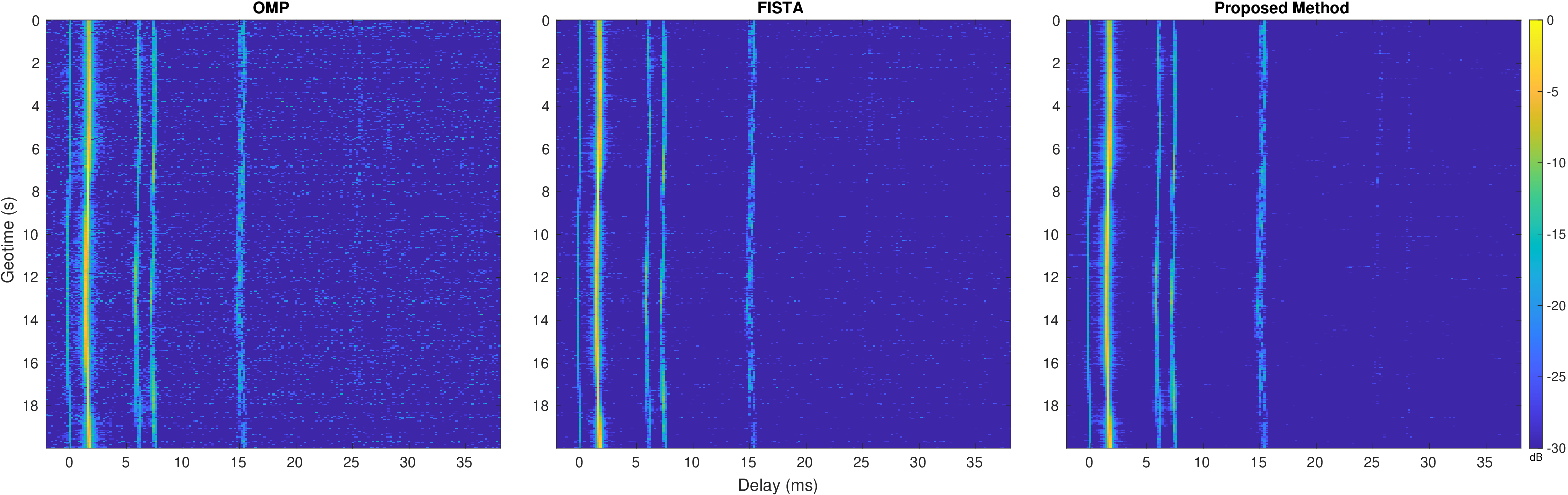}
\label{fig.cir_imp_snr_15dB_inr_40dB}}
\hfill
\subfloat[]{\includegraphics[scale = 0.17]{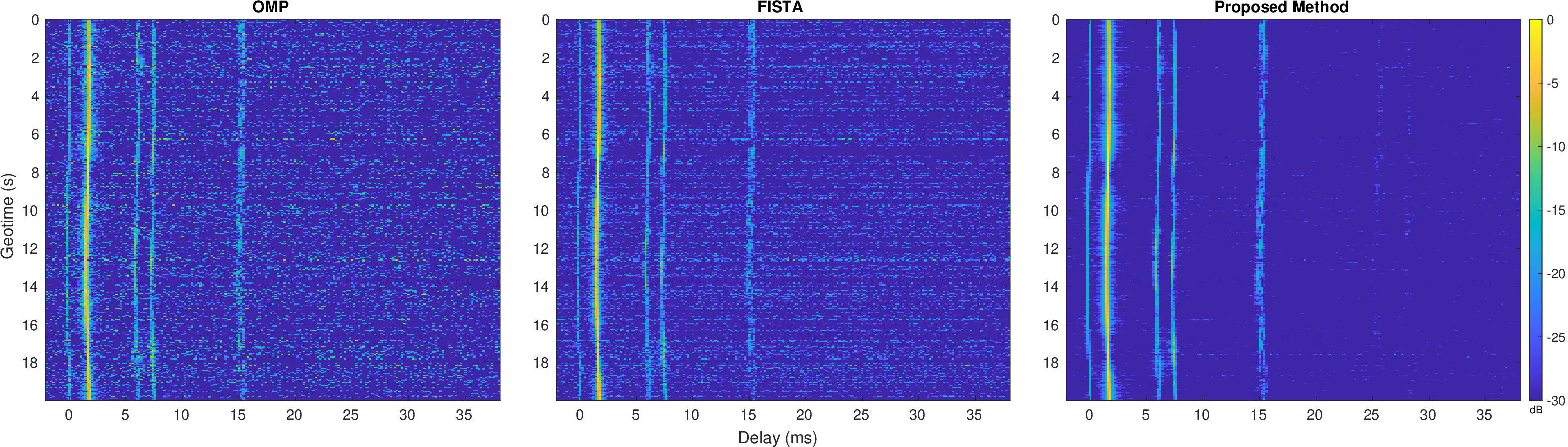}
\label{fig.cir_imp_snr_15dB_inr_50dB}}
\caption{Estimated \ac{cir} matrices under different noise environments. (a) \ac{awgn}, (b) INR = 40dB, (c) INR = 50dB.}
\label{fig.cir}
\end{figure}

\begin{figure}[!t]
\centering
\subfloat[]{\includegraphics[scale = 0.17]{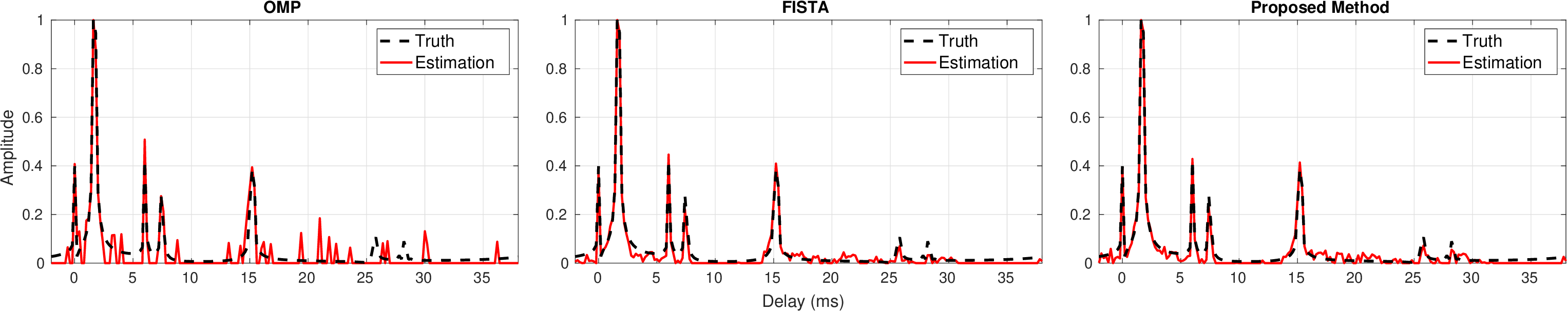}
\label{fig.sample_awgn_snr_15dB}}
\hfill
\subfloat[]{\includegraphics[scale = 0.17]{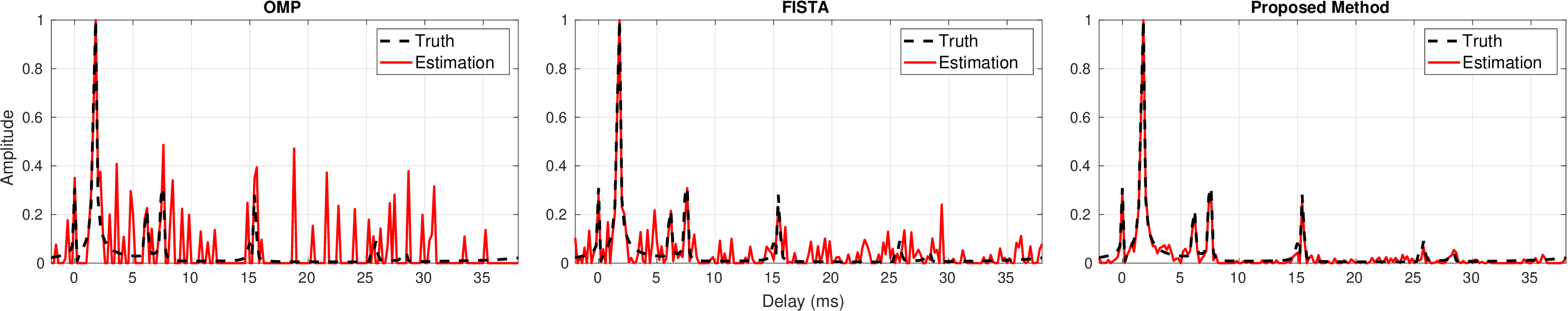}
\label{fig.sample_imp_snr_15dB_inr_40dB}}
\hfill
\subfloat[]{\includegraphics[scale = 0.17]{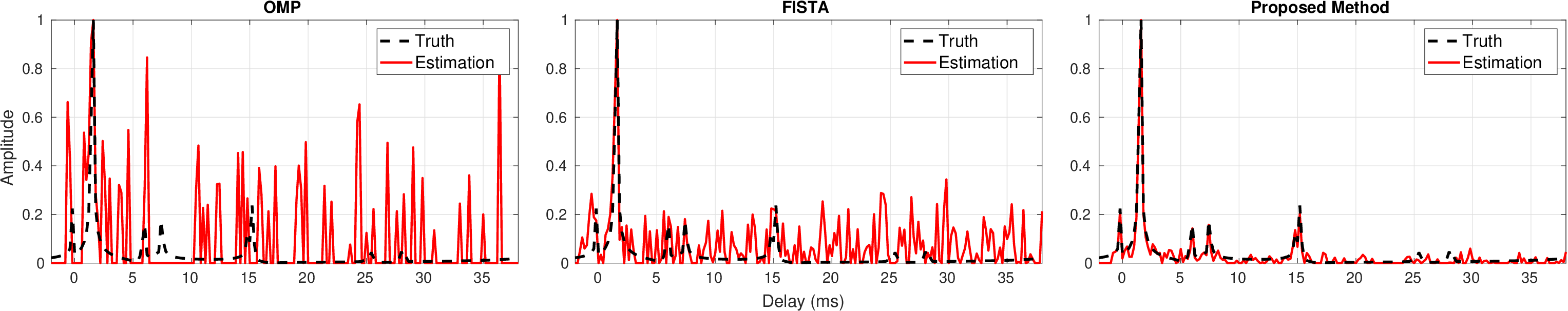}
\label{fig.sample_imp_snr_15dB_inr_50dB}}
\caption{Comparison of estimated \ac{cir} samples under different noise environments. (a) \ac{awgn}, (b) INR = 40dB, (c) INR = 50dB.}
\label{fig.cir_sample}
\end{figure}

During the simulations, the number of iterations for the \ac{omp} algorithm is set to its optimal value. The regularization parameters of the \ac{fista} and proposed algorithm are set to $\lambda=0.01\lambda_{\infty}$ and $\tau=1/(0.04\lambda_{\infty})$, respectively, where $\lambda_{\infty}=||2\mb{A}^H\mb{y}||_{\infty}$. The step-size scale factor $\eta$ in the backtracking line search is set to 1.5 for both algorithms. For the proposed algorithm, the initial penalty parameter $\rho^{(0)}$  is set to 1, and $\epsilon_{\text{abs}}=10^{-3}$ and $\epsilon_{\text{rel}}=10^{-2}$ are used as stopping criteria. 

\begin{figure}[!t]
\centering
\subfloat[]{\includegraphics[scale = 0.17]{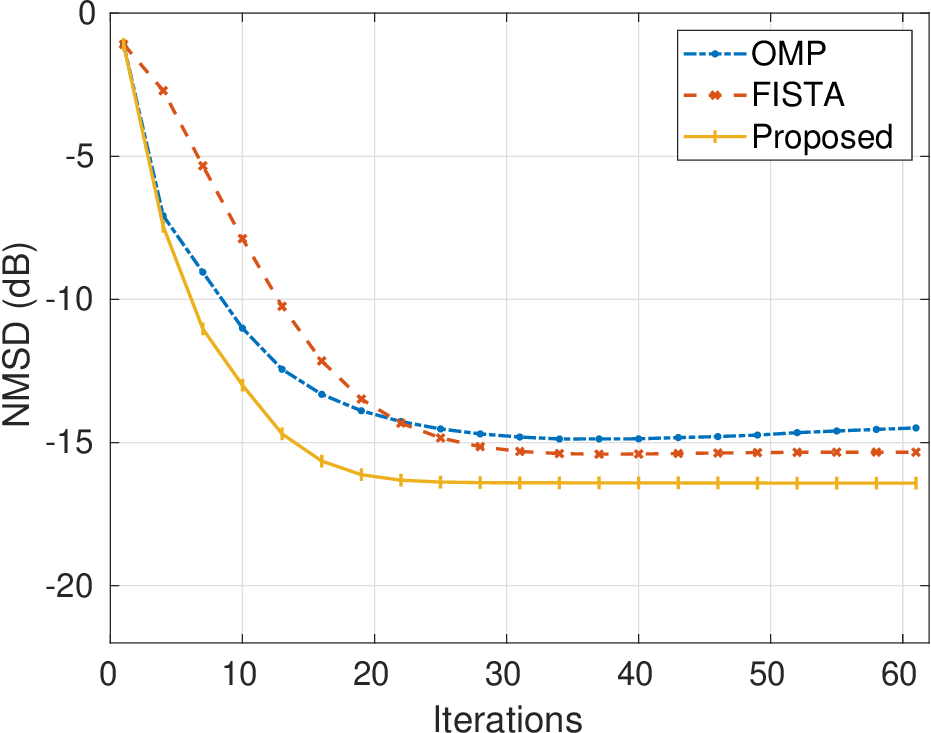}
\label{fig.mse_awgn_snr_15dB}}
\hfill
\subfloat[]{\includegraphics[scale = 0.17]{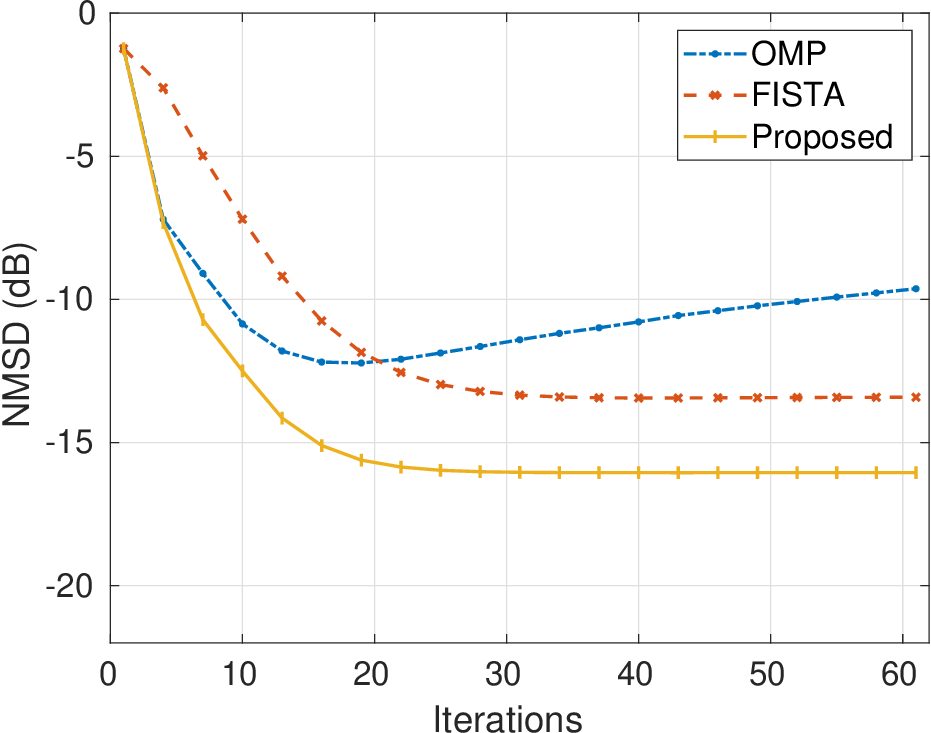}
\label{fig.mse_imp_snr_15dB_inr_40dB}}
\hfill
\subfloat[]{\includegraphics[scale = 0.17]{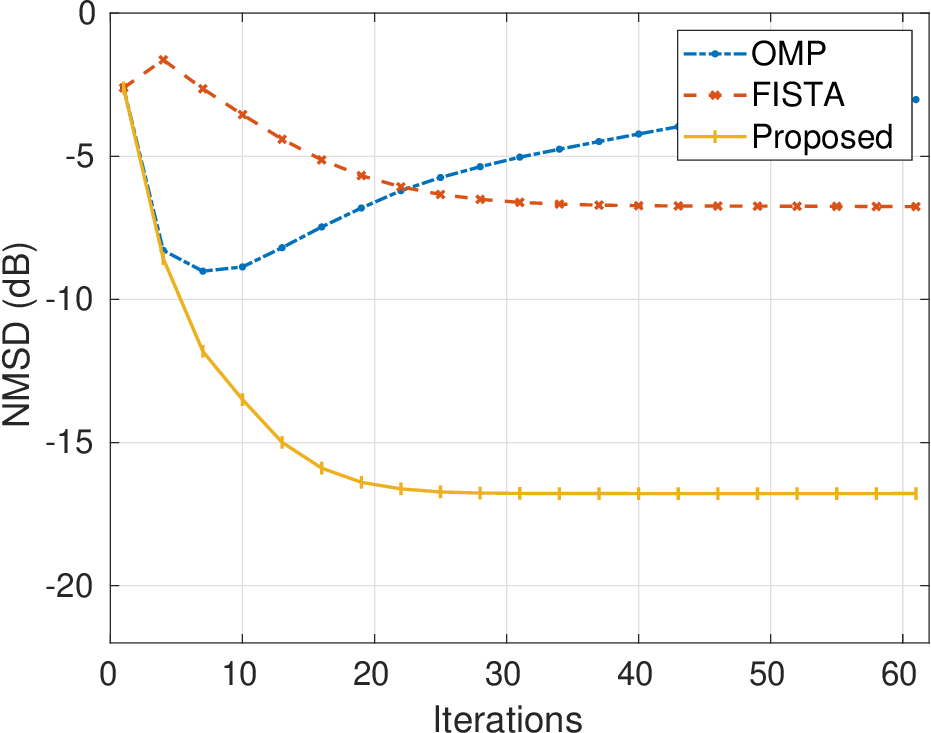}
\label{fig.mse_imp_snr_15dB_inr_50dB}}
\caption{Comparison of \ac{nmsd} curves under different noise conditions, (a) \ac{awgn}, (b) INR = 40dB, (c) INR = 50dB}
\label{fig.mse}
\end{figure}

Fig. \ref{fig.cir} shows the estimated \ac{cir} matrices by the three algorithms under different noise conditions. The corresponding estimated samples obtained from the \ac{cir} matrices are shown in Fig. \ref{fig.cir_sample}. From Fig. \ref{fig.cir} and Fig. \ref{fig.cir_sample}, one can see that all three algorithms perform similarly in the \ac{awgn} environment. However, the performance of the \ac{omp} and \ac{fista} algorithms degrades significantly in impulsive noise environments. The estimated CIR samples contain a large number of noisy taps that are supposed to be inactive. In contrast, the proposed algorithm maintains stable performance and provides accurate estimates with small errors in all three noise conditions. The \ac{nmsd} versus iteration curves are plotted in Fig. \ref{fig.mse}. The \ac{nmsd}, which measures the estimation accuracy, is defined as $\text{NMSD} = 20 \log_{10}(||\mb{x}^*- \hat{\mb{x}}||_2/||\mb{x}^*||_2))$, where $\mb{x}^*$ is the true \ac{cir} and $\hat{\mb{x}}$ is the estimated channel response. The \ac{nmsd} curves shows that the proposed algorithm achieves lowest \ac{nmsd} values of around -16dB across all three noise environments and converges within just a few tens of iterations.

\begin{figure}[!t]
\centering
\subfloat[]{\includegraphics[scale = 0.17]{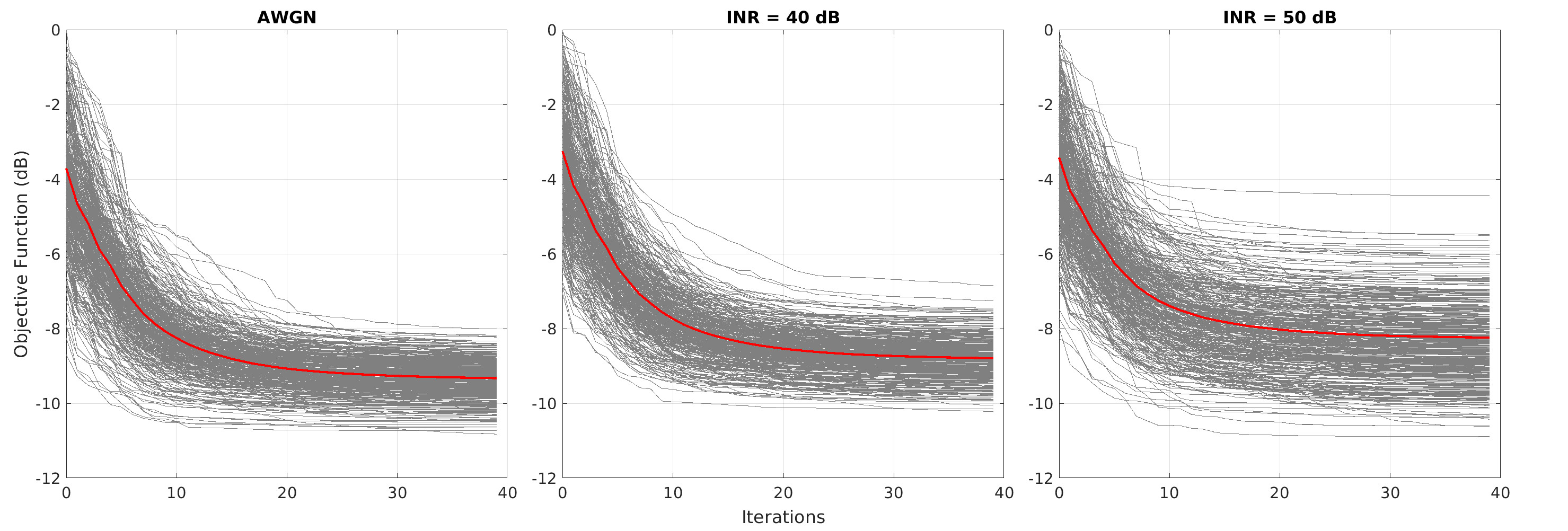}
\label{fig.admm_obj}}
\hfill
\subfloat[]{\includegraphics[scale = 0.17]{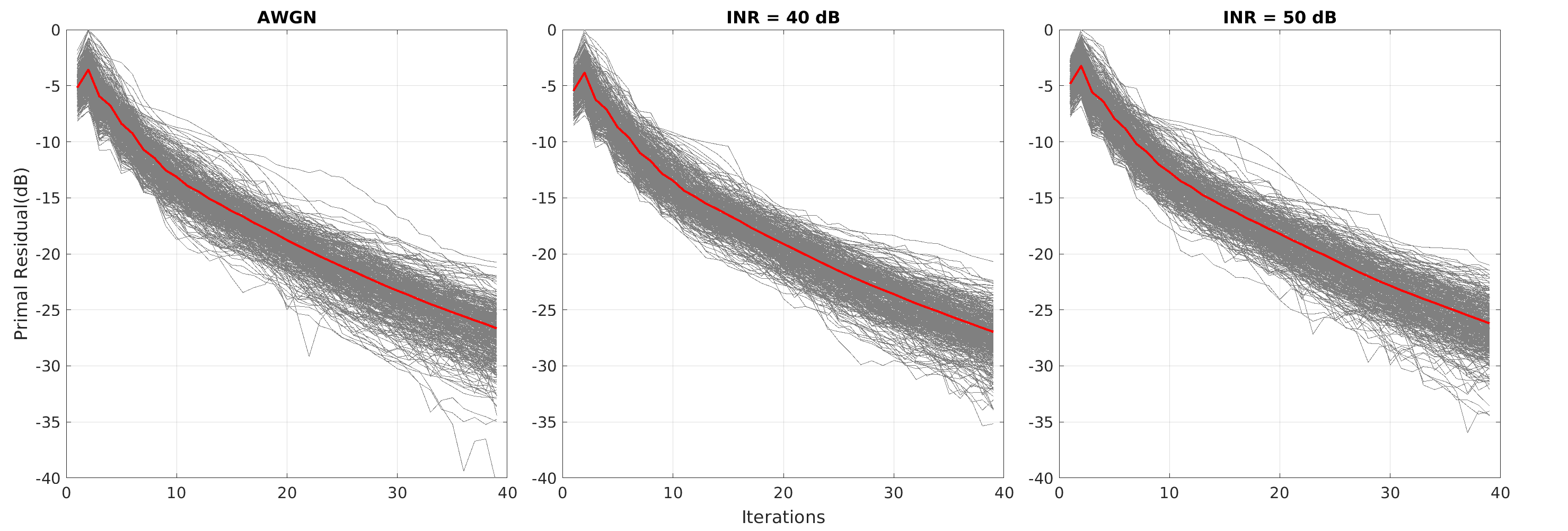}
\label{fig.admm_rnorm}}
\hfill
\subfloat[]{\includegraphics[scale = 0.17]{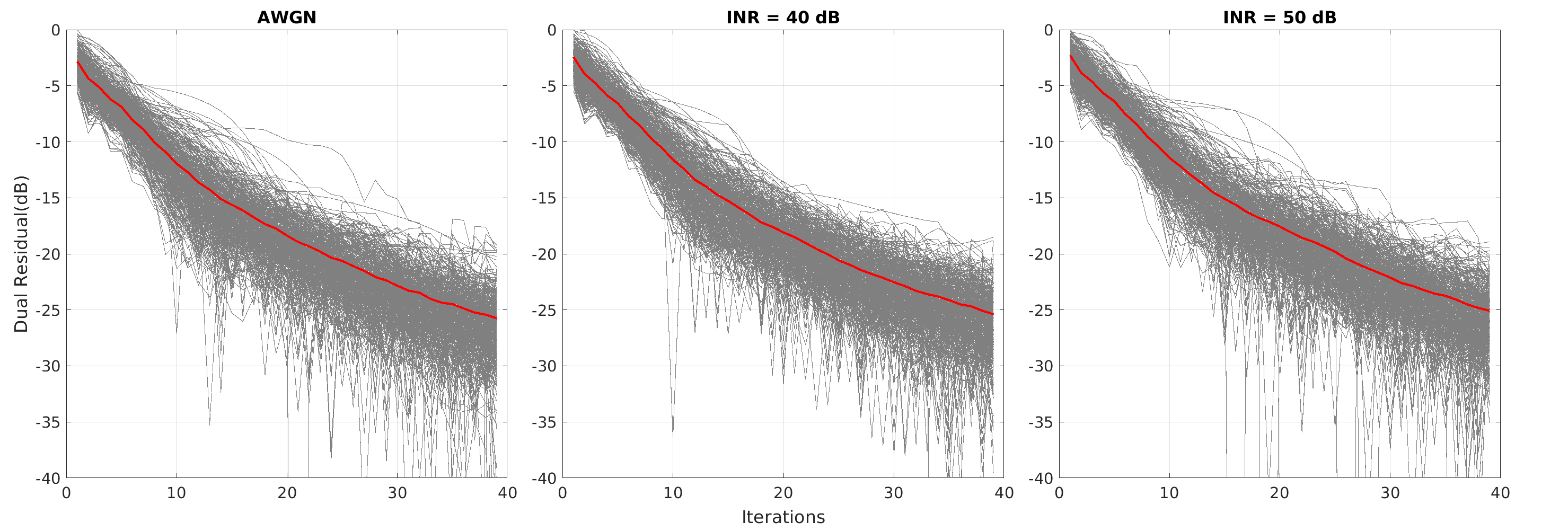}
\label{fig.admm_snorm}}
\caption{Convergence behavior of the proposed algorithm under different noise conditions. (a) objective function, (b) the norm of the primal residue, and (c) the norm of the dual residue (see \eqref{eq: sec3_rp}-\eqref{eq: sec3_stop_ccriteria}). The grey lines show the convergence behavior from single estimation process, while the red line shows the average over all samples.}
\label{fig.admm}
\end{figure}

In addition, convergence behaviors in terms of the objective function values, as well as the $\ell_2$-norm of the primal and dual residues of the proposed algorithm are illustrated in Fig. \ref{fig.admm}. We summarize the results of average number of iterations, runtime for single estimation, and final \ac{nmsd} values of three algorithms in Tab. \ref{tab. 1}. The results highlight the robustness of the proposed algorithm in impulsive noise environments.

\begin{table*}[hpt]\centering
\ra{1}
\caption{Performance comparison of three algorithms in terms of the average number of iterations, runtime, and \ac{nmsd} under different noise conditions.}
\begin{tabular}{@{}lccccccccccc@{}}\toprule
& \multicolumn{3}{c}{OMP} & \phantom{abc}& \multicolumn{3}{c}{FISTA} &
\phantom{abc} & \multicolumn{3}{c}{Proposed Method}\\
\cmidrule{2-4} \cmidrule{6-8} \cmidrule{10-12}
$\sigma_{I}^2/\sigma_{W}^2$ & $-\inf$ & $40$dB & $50$dB && $-\inf$ & $40$dB & $50$dB && -$\inf$ & $40$dB & $50$dB\\ \midrule
\# Iter & 33.9 & 43.1 & 48.2 && 33.9 & 33.1 & 32.6 && 34.7 & 33.9 & 34.0\\
Runtime (s) & 0.024 & 0.052 & 0.052 && 0.026 & 0.024 & 0.026 && 0.051& 0.050 & 0.050\\
NMSD (dB) & -14.77 & -12.18 & -9.01 && -15.37 & -13.39 & -6.65 && -16.42 & -16.05 & -16.78\\
\bottomrule
\end{tabular}
\label{tab. 1}
\end{table*}

%% file: sections/4_conclusion.tex
\section{Conclusions}
This work developed a robust \ac{admm}-based channel estimation algorithm available for impulsive noise environments. The proposed algorithm is low in complexity and easy to be implemented. We evaluate the performance of the proposed algorithm in solving the sparse channel estimation problem and compare it with popular \ac{omp} and \ac{fista} algorithms. The results highlight the effectiveness and robustness of the proposed algorithm, making it a practical choice for channel estimation in underwater acoustic communication systems.

%% file: sections/5_acknowledgements.tex
\section{Acknowledgement}
This research has been funded in part by the National Natural Science Foundation of China (Project No. 62171369), and in part by the Key Program of the National Natural Science Foundation of China (Grant No. 52231013). Additionally, this work is also sponsored by the China Scholarship Council.

%% file: acronyms.tex
\begin{acronym}
  \acro{ser}[SER]{symbol error rate}
  \acro{uacs}[UACs]{underwater acoustic channels}
  \acro{uac}[UAC]{underwater acoustic channel}
  \acro{uwa}[UWA]{underwater acoustic}
  \acro{ti}[TI]{time-invariant}
  \acro{dd}[DD]{delay-Doppler}
  \acro{ici}[ICI]{intercarrier interference}
  \acro{cir}[CIR]{channel impulse response}
  \acro{cp}[CP]{cyclic prefix}
  \acro{dft}[DFT]{discrete Fourier transform}
  \acro{ofdm}[OFDM]{Orthogonal Frequency Division Multiplexing }
  \acro{ipm}[IPM]{interior-point method}
  \acro{admm}[ADMM]{alternating direction method of multipliers}
  \acro{pgm}[PGM]{proximal gradient method}
  \acro{apg}[APG]{accelerated proximal gradient}
  \acro{alf}[ALF]{augmented Lagrangian function}
  \acro{alm}[ALM]{augmented Lagrangian method}
  \acro{omp}[OMP]{orthogonal matching pursuit}
  \acro{fista}[FISTA]{fast iterative shrinkage-thresholding algorithm}
  
  \acro{snr}[SNR]{signal-to-noise ratio}
  \acro{nmsd}[NMSD]{normalized mean-square deviation}
  \acro{wgn}[WGN]{white Gaussian noise}
  \acro{gmn}[GMN]{Gaussian mixture noise}
  \acro{cs}[CS]{compressed sensing}
  \acro{rip}[RIP]{restricted isometry property}
  \acro{caf}[CAF]{cross-ambiguity function}
  \acro{mmse}[MMSE]{minimum mean squared error }
  \acro{ber}[BER]{bit error rate}
  \acro{ser}[SER]{symbol error rate}
  \acro{prbs}[PRBS]{pseudo-random binary sequence}
  \acro{bpsk}[BPSK]{Binary Phase-shift keying}
  \acro{qpsk}[QPSK]{Quadrature Phase Shift Keying}
  \acro{srrc}[SRRC]{Square root raised cosine}
  \acro{caf}[CAF]{cross-ambiguity function}
  \acro{inr}[INR]{interference-to-noise ratio}
  \acro{awgn}[AWGN]{additive white Gaussian noise}
  \acro{sinr}[SINR]{signal-to-interference-plus-noise ratio}

  \acro{ls}[LS]{Least Squares}
  \acro{pn}[PN]{pseudorandom noise}
  \acro{mimo}[MIMO]{Multiple-Input and Multiple-Output}
  \acro{uwa}[UWA]{Underwater acoustic}
  \acro{rt}[RT]{Real Time}
  \acro{ric}[RIC]{RAN Intelligent Controller}
  \acro{mac}[MAC]{Medium Access Control}
  \acro{bs}[BS]{Base Station}
  \acro{ue}[UE]{User Equipment}
  \acro{cqi}[CQI]{Channel Quality Indicator}
  \acro{mcs}[MCS]{Modulation and Coding Scheme}
  \acro{pusch}[PUSCH]{Physical Uplink Shared Channel}
  \acro{pucch}[PUCCH]{Physical Uplink Control Channel}
  \acro{usrp}[USRP]{Universal Software Radio Peripheral}  \acro{vm}[VM]{Virtual Machine}
  \acro{voip}[VoIP]{Voice over IP}
  \acro{dos}[DoS]{Denial-of-Service}
  \acro{tcp}[TCP]{Transmission Control Protocol}
  \acro{qos}[QoS]{Quality of Service}
  \acro{udp}[UDP]{User Datagram Protocol}
  \acro{svm}[SVM]{Support Vector Machine}
  \acro{knn}[k-NN]{k-Nearest Neighbors}
  \acro{adaboost}[AdaBoost]{Adaptive Boosting}
  \acro{mlp}[MLP]{Multilayer Perceptron}
  \acro{ids}[IDS]{Intrusion Detection System}
  \acro{du}[DU]{Distributed Unit}
  \acro{cu}[CU]{Centralized Unit}
  \acro{rrc}[RRC]{Radio Resource Control}
  \acro{dpi}[DPI]{Deep Packet Inspection}
\end{acronym}

%% file: main.bbl
\begin{thebibliography}{10}
\providecommand{\url}[1]{#1}
\csname url@samestyle\endcsname
\providecommand{\newblock}{\relax}
\providecommand{\bibinfo}[2]{#2}
\providecommand{\BIBentrySTDinterwordspacing}{\spaceskip=0pt\relax}
\providecommand{\BIBentryALTinterwordstretchfactor}{4}
\providecommand{\BIBentryALTinterwordspacing}{\spaceskip=\fontdimen2\font plus
\BIBentryALTinterwordstretchfactor\fontdimen3\font minus
  \fontdimen4\font\relax}
\providecommand{\BIBforeignlanguage}[2]{{%
\expandafter\ifx\csname l@#1\endcsname\relax
\typeout{** WARNING: IEEEtran.bst: No hyphenation pattern has been}%
\typeout{** loaded for the language `#1'. Using the pattern for}%
\typeout{** the default language instead.}%
\else
\language=\csname l@#1\endcsname
\fi
#2}}
\providecommand{\BIBdecl}{\relax}
\BIBdecl

\bibitem{wenRobustSparseRecovery2017}
F.~Wen, P.~Liu, Y.~Liu, R.~C. Qiu, and W.~Yu, ``Robust {{Sparse Recovery}} in
  {{Impulsive Noise}} via $\ell_p$-$\ell_1$ {{Optimization}},'' \emph{{IEEE}
  Trans. Signal Process.}, vol.~65, no.~1, pp. 105--118, Jan. 2017.

\bibitem{boydDistributedOptimizationStatistical2010a}
S.~Boyd, ``Distributed {{Optimization}} and {{Statistical Learning}} via the
  {{Alternating Direction Method}} of {{Multipliers}},'' \emph{FNT in Machine
  Learning}, vol.~3, no.~1, pp. 1--122, 2010.

\bibitem{zhangOnlineProximalADMMTimeVarying2021}
Y.~Zhang, E.~Dall'Anese, and M.~Hong, ``Online {{Proximal-ADMM}} for
  {{Time-Varying Constrained Convex Optimization}},'' \emph{IEEE Transactions
  on Signal and Information Processing over Networks}, vol.~7, pp. 144--155,
  2021.

\bibitem{parikhProximalAlgorithms2014}
N.~Parikh and S.~Boyd, ``Proximal {{Algorithms}},'' \emph{OPT}, vol.~1, no.~3,
  pp. 127--239, Jan. 2014.

\bibitem{hagerInexactAlternatingDirection2019}
W.~W. Hager and H.~Zhang, ``Inexact alternating direction methods of
  multipliers for separable convex optimization,'' \emph{Computational
  Optimization and Applications}, vol.~73, no.~1, pp. 201--235, May 2019.

\bibitem{daspremontAccelerationMethods2021}
A.~{d'Aspremont}, D.~Scieur, and A.~Taylor, ``Acceleration {{Methods}},''
  \emph{Foundations and Trends\textregistered{} in Optimization}, vol.~5, no.
  1-2, pp. 1--245, 2021.

\bibitem{wohlbergADMMPenaltyParameter2017}
B.~Wohlberg, ``{{ADMM Penalty Parameter Selection}} by {{Residual
  Balancing}},'' Apr. 2017.

\bibitem{caiOrthogonalMatchingPursuit2011}
T.~T. Cai and L.~Wang, ``Orthogonal {{Matching Pursuit}} for {{Sparse Signal
  Recovery With Noise}},'' \emph{IEEE Transactions on Information Theory},
  vol.~57, no.~7, pp. 4680--4688, Jul. 2011.

\bibitem{beckFastIterativeShrinkageThresholding2009a}
A.~Beck and M.~Teboulle, ``A {{Fast Iterative Shrinkage-Thresholding
  Algorithm}} for {{Linear Inverse Problems}},'' \emph{SIAM Journal on Imaging
  Sciences}, vol.~2, no.~1, pp. 183--202, Jan. 2009.

\bibitem{qarabaqiStatisticalCharacterizationClass2014}
P.~Qarabaqi, ``Statistical characterization of a class of underwater acoustic
  communication channels,'' Ph.D. dissertation, Northeastern University, 2014.

\end{thebibliography}
